\documentclass[useAMS, usenatbib]{mn2e}
\usepackage{graphicx}
\usepackage{paralist}
\usepackage{url}
\usepackage{color, soul}	%for highlighting
\usepackage{xcolor}

\begin{document}

\title[Constraining the EoR with the variance statistic]{Constraining the epoch of reionization with the variance statistic: simulations of the LOFAR case}

\author[A. H. Patil et al.]
{Ajinkya H. Patil,$^{1}$\thanks{E-mail: patil@astro.rug.nl}
Saleem Zaroubi,$^{1}$ Emma Chapman,$^{2}$ Vibor Jeli\'{c},$^{1,3}$ 
\newauthor Geraint Harker,$^{4}$ Filipe B. Abdalla,$^{2}$ Khan M. B. Asad,$^{1}$ Gianni Bernardi,$^{5}$ 
\newauthor Michiel A. Brentjens,$^{3}$ A. G. de Bruyn,$^{1,3}$ Sander Bus,$^{1}$ Benedetta Ciardi,$^{6}$
\newauthor Soobash Daiboo,$^{1}$ Elizabeth R. Fernandez,$^{1}$ Abhik Ghosh,$^{1}$ Hannes Jensen,$^{7}$
\newauthor Sanaz Kazemi,$^{1}$ L\'{e}on V. E. Koopmans,$^{1}$ Panagiotis Labropoulos,$^{3}$ Maaijke Mevius,$^{1,3}$
\newauthor Oscar Martinez,$^{1}$ Garrelt Mellema,$^{7}$ Andre. R. Offringa,$^{1,8}$ Vishhambhar N. Pandey,$^{1,3}$
\newauthor Joop Schaye,$^{9}$ Rajat M. Thomas,$^{1}$ Harish K. Vedantham,$^{1}$ Vamsikrishna Veligatla,$^{1}$
\newauthor Stefan J. Wijnholds$^{3}$ and Sarod Yatawatta$^{1,3}$\\
$^{1}$Kapteyn Astronomical Institute, University of Groningen, PO Box 800, 9700AV Groningen, the Netherlands\\
$^{2}$Department of Physics \& Astronomy, University College London, Gower Street, London WC1E 6BT\\
$^{3}$ASTRON, PO Box 2, NL-7990AA Dwingeloo, the Netherlands\\
$^{4}$Center for Astrophysics and Space Astronomy, University of Colorado, 389 UCB, Boulder, Colorado 80309-0389, USA\\
$^{5}$SKA SA, 3rd Floor, The Park, Park Road, Pinelands, 7405, South Africa\\
$^{6}$Max-Planck Institute for Astrophysics, Karl-Schwarzschild-Strasse 1, D-85748 Garching bei M\"unchen, Germany\\
$^{7}$Department of Astronomy and Oskar Klein Centre, Stockholm University, AlbaNova, SE-10691 Stockholm, Sweden\\
$^{8}$Mount Stromlo Observatory, RSAA, Cotter Road, Weston Creek ACT 2611, Australia\\
$^{9}$Leiden Observatory, Leiden University, PO Box 9513, NL-2300 RA Leiden, the Netherlands}

\date{}
\pagerange{}
\pubyear{}
\maketitle{}

\begin{abstract}
Several experiments are underway to detect the cosmic redshifted 21-cm signal from neutral hydrogen from the Epoch of Reionization (EoR). Due to their very low signal-to-noise ratio, these observations aim for a statistical detection of the signal by measuring its power spectrum. We investigate the extraction of the variance of the signal as a first step towards detecting and constraining the global history of the EoR. Signal variance is the integral of the signal's power spectrum, and it is expected to be measured with a high significance. We demonstrate this through results from a simulation and parameter estimation pipeline developed for the Low Frequency Array (LOFAR)-EoR experiment. We show that LOFAR should be able to detect the EoR in 600 hours of integration using the variance statistic. Additionally, the redshift ($z_r$) and duration ($\Delta z$) of reionization can be constrained assuming a parametrization. We use an EoR simulation of $z_r = 7.68$ and $\Delta z = 0.43$ to test the pipeline. We are able to detect the simulated signal with a significance of 4 standard deviations and extract the EoR parameters as  $z_r = 7.72^{+0.37}_{-0.18}$ and $\Delta z = 0.53^{+0.12}_{-0.23}$ in 600 hours, assuming that systematic errors can be adequately controlled. We further show that the significance of detection and constraints on EoR parameters can be improved by measuring the cross-variance of the signal by cross-correlating consecutive redshift bins.
\end{abstract}


\section{Introduction}
Advances in observational cosmology over the past century have made it possible to look very far out into the Universe. However, there still remains a big observational gap between the Cosmic Microwave Background (CMB) ($z \approx 1100$) and the low-redshift Universe ($z<6$). An important global transition is expected to have occurred towards the end of this era, called the Epoch of Reionization (EoR). The first stars and galaxies formed during this epoch, and hydrogen in the Universe was reionized from their radiation after having been neutral for about 400 Myr. 

Unfortunately, a dearth of observations makes the EoR a poorly constrained epoch. Current constraints are based on indirect observations of the high-redshift intergalactic medium (IGM), namely, quasar spectra \citep{Fan2003, Fan2006}, CMB polarization anisotropy \citep[e.g.][]{Hinshaw2013}, the kinetic Sunyaev-Zel'dovich effect \citep{Zahn2012}, IGM temperature measurements \citep{Theuns2002, Bolton2010}, high-redshift galaxy surveys \citep[e.g.][]{Finkelstein2012}, high redshift gamma ray bursts \citep{Wang2013} and Lyman break galaxies \citep{Pentericci2011, Ono2012, Schenker2012}.  However, redshifted 21-cm emission from neutral hydrogen has the potential to directly probe the IGM and hence the process of reionization. Therefore, many ongoing experiments aim to observe the EoR with low-frequency radio telescopes such as the Low Frequency Array (LOFAR) \citep{vanHaarlem2013}, the Murchison Widefield Array (MWA) \citep{Tingay2013}, the Precision Array to Probe the Epoch of Reionization (PAPER) \citep{Parsons2010} and the Giant Meterwave Radio Telescope (GMRT) \citep{Pen2008}.

Detection of the EoR signal, i.e. the redshifted 21-cm signal from the era of reionization, is very challenging even with the new generation of radio telescopes. This is because in the redshift range of 6 to 10, the expected signal is only about 10 mK (at a resolution of 3 arcmin), whereas, the Galactic and extragalactic foregrounds are about 1 K \citep{Bernardi2009, Bernardi2010}. Moreover, even if the foregrounds would be perfectly removed, after hundreds of hours of integration, the system noise would still be an order of magnitude larger than the signal. Therefore, the current experiments aim for a statistical detection of the EoR rather than mapping the neutral hydrogen. This requires development of the  statistical techniques to estimate the reionization parameters from noisy data. A commonly studied technique is power spectrum analysis \citep[e.g.][]{Morales2004, McQuinn2006, Bowman2006, Harker2010, Beardsley2013}. Another possible statistic is the signal variance, which is the integral of signal's power spectrum. The variance (or root mean square) statistics of the EoR signal has been studied theoretically by \cite{Iliev2008, Jelic2008, Thomas2009, Harker2009, Bittner2011, Watkinson2013}. 

In this paper, we investigate the measurement of signal variance with LOFAR considering various instrumental parameters.  We use the variance as a quantitative measure to constrain the global history of reionization in early stage EoR experiments. We have developed a simulation pipeline to test the variance statistic in the case of the LOFAR-EoR experiment. The pipeline generates mock observations by simulating the cosmic signal, foregrounds and noise. The pipeline also incorporates measurement of the EoR parameters, namely, the redshift and duration of the EoR.

The paper is organized as follows: in Section 2, we discuss our parametrization of the variance of the EoR signal as a function of redshift. In Section 3, we describe the simulation pipeline we have developed. The measurement of the signal variance and parameter estimation is discussed in Section 4. Here we also discuss the advantages of measuring the cross-variance of the signal by cross-correlation consecutive frequency bins. We show the results and demonstrate that LOFAR should be able to constrain the EoR in 600 h in Section 5, before summarizing our conclusions in Section 6.

\section{Parametrization}

The observable quantity of the redshifted 21-cm emission is the differential brightness temperature $\delta T_{\rmn{b}}$ i.e. the contrast between the 21-cm brightness temperature and the background CMB temperature $T_{\rmn{CMB}}$. At a given position in sky, $\delta T_{\rmn{b}}$ is given by \citep{Field1959, Madau1997, Furlanetto2006}
\begin{eqnarray}
\delta T_{\rmn{b}} \approx 9 \; x_{\rmn{HI}} (1+\delta) (1+z)^{\frac{1}{2}} \left[1 - \frac{T_{\rmn{CMB}}(z)}{T_{\rmn{S}}}\right] \nonumber \\
\times \left[\frac{H(z)/(1+z)}{dv_{\parallel}/dr_{\parallel}}\right] \rmn{mK},
\end{eqnarray}
where $\delta$ is the cosmological mass density contrast, $x_{\rmn{HI}}$ is the neutral hydrogen fraction, $T_{\rmn{S}}$ is the spin temperature, $H(z)$ is Hubble parameter and $dv_{\parallel}/dr_{\parallel}$ is gradient of the proper velocity along the line of sight. Whenever we mention the EoR signal, we refer to the differential brightness temperature of the 21-cm radiation from reionization.

An interferometer can measure spatial fluctuations of $\delta T_{\rmn{b}}$ as a function of frequency, or equivalently of cosmic redshift. The spatial fluctuations at a given redshift can be characterized by the power spectrum $P[\mathbfit{k}]$ as
\begin{equation}
P[k] = \langle \delta T_b [\mathbfit{k}] \delta T_b ^*[\mathbfit{k}] \rangle _{\left| \mathbfit{k} \right| = k}
\end{equation}
where $\delta T_{\rmn{b}}$ is measured at discrete values of wavenumber $\mathbf{k}$. The variance of the signal is the average over $k$ as given by
\begin{equation}
\rmn{Var}(\delta T_b) = \langle P[k] \rangle.
\end{equation} 
Our interest here lies in measuring the variance of the signal and its evolution with redshift.

Fig. 1 shows the evolution of the signal variance as predicted by the simulation code \textsc{21cmFAST} \citep{Mesinger2011}. At the highest redshifts, the Universe
is mostly neutral ($x_{\rmn{HI}} \approx 1$), hence $\delta T_{\rmn{b}}$ is driven by the cosmological density fluctuations $\delta$. The density fluctuations grow with time to form the first ionizing sources, which then start to reionize their surrounding regions. This patchy nature of reionization leads to a rise in the variance of $\delta T_{\rmn{b}}$. The variance reaches its peak when approximately half of the Universe is ionized, but decreases thereafter. Eventually, it reaches zero as the entire Universe is reionized. The different curves in Fig. 1 are for different spatial resolutions and show that the observed variance depends on the resolution, or equivalently on the range of wavenumbers measured in the observed volume. Also, the higher the resolution, the earlier the variance peaks \citep{Iliev2008}. This is because higher resolution data are sensitive to smaller scale structures.  

\begin{figure}
\centering
\includegraphics{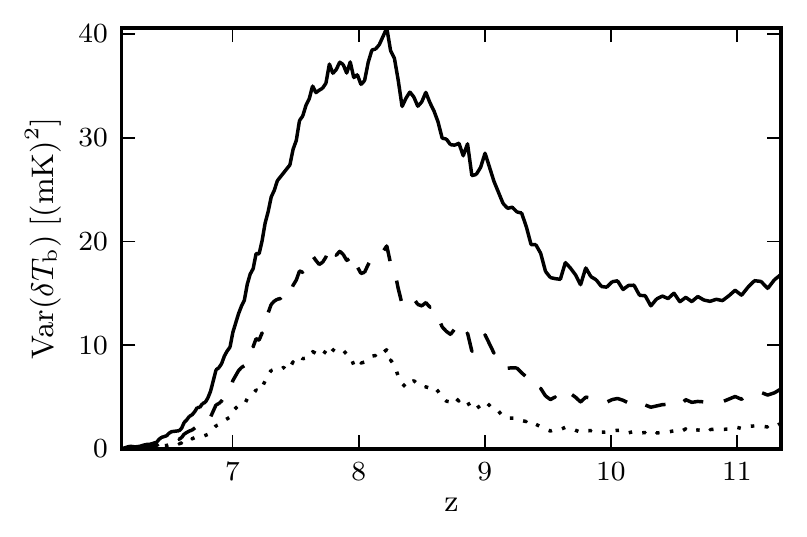}
\caption{Evolution of the variance of the epoch of reionization signal i.e. the 21-cm differential brightness temperature ($\delta T_{\rmn{b}}$) with redshift, as predicted by a simulation with \textsc{21cmFAST}. From top to bottom the curves correspond to decreasing spatial resolutions of 3 (resolution of the LOFAR core i.e. the central 2 km of the array at 150 MHz, $z \approx 8.5$), 7 and 12 arcmin. The resolution is expressed as the full width half maximum (FWHM) of the Gaussian kernel that was used to smooth the simulation maps.}
\end{figure}	

In order to learn about the process of reionization from variance measurements, we need a parametric model which describes the variance of $\delta T_{\rmn{b}}$ in terms of the EoR model parameters. In this paper, we assume a model which enables us to constrain two important EoR parameters: the redshift of reionization $z_\rmn{r}$, defined as the redshift at which the variance of $\delta T_{\rmn{b}}$ is maximum, and the duration of reionization $\Delta z$. The model is given by
\begin{equation}
\mathrm{Var} ( \delta T_{\rmn{b}} ) = A \; f(z)\left( \frac{z}{z_{0}} \right) ^{\beta},
\end{equation}
where $A$ is the scaling amplitude, $\beta$ $(<0)$ is the index of the (decaying) power law that the variance asymptotes to at high redshift and $z_0$ is the redshift which defines the regime $z \gg z_0$ in which the power law becomes dominant. The model is inspired by the fact that at high redshift, $\delta T_{\rmn{b}}$ is driven by $(1+\delta)$, which linear perturbation theory predicts to evolve as a power law. The function $f(z)$ describes the low-redshift behaviour of the signal and is defined as
\begin{equation}
f(z) = 1 + \tanh \left( \frac{z - z_{0}}{\Delta z} \right).
\end{equation}

The redshift of reionization $z_r$ is the redshift at which the variance reaches its maximum. Therefore, it is computed using the condition
\begin{equation}
\frac{d\mathrm{Var}(\delta T_{\rmn{b}})}{dz} \Big|_{z_r} = 0.
\end{equation}
We translate the parameter $z_0$ to $z_r$ by computing the difference $z_{bias}$ between the two and then correcting for it as
\begin{equation}
z_r = z_0 + z_{bias}.
\end{equation}

\begin{figure}
\centering
\includegraphics{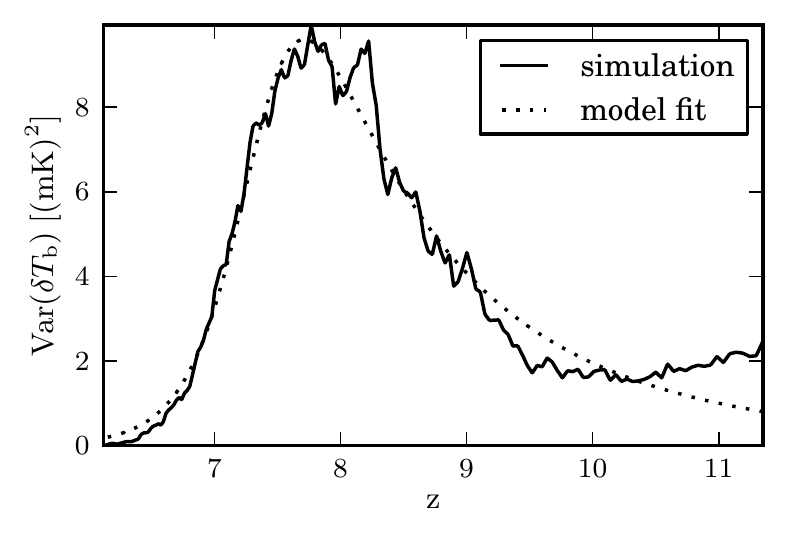}
\caption{Fit of the model described by equations (4) and (5) to the signal variance predicted by the simulation at 12 arcmin resolution. The model describes the curve well except at $z > 9$.}
\end{figure}

Fig. 2 shows a model fit to the signal variance from a simulation. It can be seen that the model describes the simulation results well, except for the dip in the variance at high redshifts $(z \approx 10)$. Such a decrease in the variance is expected to occur at the beginning of reionization \citep{Iliev2012}. The first objects form in density peaks and reionize their surrounding regions, which appear as holes in neutral hydrogen maps. These holes reduce the signal variance contributed by the corresponding density peaks. However, when many ionizing objects start to form, the variance is driven by the distribution of $x_{\rmn{HI}}$ rather than by the density fluctuations. Therefore, the variance increases after the initial dip. 

The small decrease in the variance at the beginning of reionization is not described by our parametrization. However, the LOFAR system noise increases rapidly at lower frequencies (higher redshifts), making the fitting performance by models which would incorporate this feature indistinguishable.

\section{Simulation and signal extraction pipeline}
We generated mock observational data sets by adding simulations of the cosmological signal, foregrounds and noise. A data cube consisted of 170 frequency maps between 115 and 199.5 MHz (i.e. $z$ = 6 to 11.4) at an interval of 0.5 MHz. Each frequency map initially represented a 10\degr $\times$ 10\degr window with 1.17 arcmin resolution but was later corrected for the LOFAR field of view as will be discussed in Section 3.3. The important blocks of the simulation and signal extraction pipeline are described in the following subsections (please see Fig. 3 for a block diagram of the pipeline.)

\begin{figure}
\centering
\includegraphics{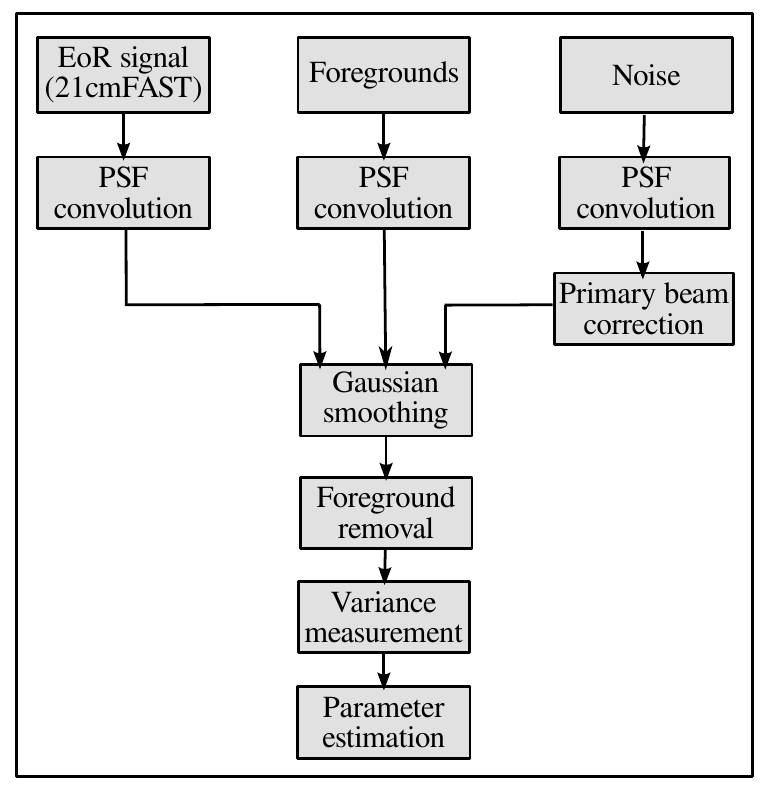}
\caption{Block diagram of the end-to-end simulation and analysis pipeline.}
\end{figure}

\subsection{The signal}
Cosmological simulations of size larger than 600 comoving Mpc are required to simulate the field of view of LOFAR. Full radiative transfer simulations on such large scales are computationally expensive. Instead, we used the semi-analytic code \textsc{21cmFAST} \citep{Mesinger2011, Mesinger2007} to simulate the EoR signal. \textsc{21cmFAST} treats physical processes with approximate methods, but on scales larger than 1 Mpc its results are in good agreement with hydrodynamical simulations \citep{Mesinger2011}. The cosmological simulation used here is the same as in Chapman et al. (2012). The simulation was initialized with $1800^3$ dark matter particles at $z = 300$. The code evolves the initial density and velocity fields to the redshifts of the EoR using linear perturbation theory. The velocity field used to perturb the initial conditions and the evolved simulation boxes were formed on a coarser grid of $450^3$ and then interpolated up to $512^3$. \textsc{21cmFAST} uses the excursion set formalism to form dark matter haloes. We define the threshold for haloes contributing ionizing photons to be $10^9 \rmn{M_{\odot}}$. Once the evolved density, velocity and ionization fields have been obtained, the code computes the $\delta T_{\rmn{b}}$ box at each redshift based on equation (1). Redshift space distortions were taken into account in our run, but we neglected spin temperature fluctuations by assuming $T_{\rmn{S}} \gg T_{\rmn{CMB}}$, i.e. the neutral gas has been heated well above the CMB for redshifts 6 to 12 \citep{Pritchard2008}. We combined the $\delta T_{\rmn{b}}$ boxes at different redshifts using the method described by \cite{Thomas2009} to form an observational cube. An observational cube represents the 2D position on the sky and the third dimension corresponds to observation frequency or redshift.

\subsection{Foregrounds}
We used the simulations by \cite{Jelic2008, Jelic2010} to model the foreground contamination. These simulations consider the following contributions:

\begin{enumerate}
\item Galactic diffuse synchrotron emission (GDSE) due to the interaction of cosmic ray electrons with the galactic magnetic field. The GDSE is modelled as a power law  as a function of frequency with a spectral index of $-2.55 \pm 0.1$ \citep{Shaver1999}. The intensity and the spectral power law index of the GDSE are spatially modelled as Gaussian random fields. The power spectrum of these fields is assumed to be a power law with 2D index of $-2.7$. The mean brightness temperature at 120 MHz is 253 K, with a standard deviation of 1.3 K.

\item Galactic localized synchrotron emission from supernova remnants (SNRs). Eight SNRs are placed randomly in the 10\degr $\times$ 10\degr observational window. In order to model the extended nature of SNRs, they are modelled to be extended discs. Their angular size, flux density and spectral index are randomly chosen from the Green (2006) catalogue of the observed radio SNRs.

The combined Galactic diffuse and localized synchrotron emission is the dominant component ($\sim$70 per cent) of the foregrounds at 100-200 MHz.

\item Galactic diffuse free-free emission due to bremsstrahlung radiation from diffuse ionized gas. It is modelled in a similar manner as the GDSE but the frequency spectral index is fixed to -2.15 across the map. It contributes $\sim$1 per cent of the total foreground emission.

\item Unresolved extragalactic sources such as radio galaxies and clusters, contributing $\sim$27 per cent of the foreground emission. The simulated radio galaxies have power law spectra and random walk based clustering. The radio clusters have spectral indices of about -3 and are based on the cluster catalogue from the Virgo Consortium\footnote{http://www.mpa-garching.mpg.de/galform/virgo/hubble/}.
\end{enumerate}

We assume that calibration would remove the point sources brighter than 0.1 mJy, hence these sources are not included in the foreground simulations \mbox{\citep{Jelic2008}.}

\subsection{Instrumental response and noise}
Unlike the EoR and foreground simulations, an interferometer does not directly map the surface brightness distribution in the sky. Instead, it measures correlations of electric fields between pairs of interferometric elements (LOFAR stations). These correlations are called visibilities. A visibility $V(u_k, v_k)$ probes a certain spatial scale of the sky brightness distribution corresponding to the baseline $(u_k, v_k)$ between a pair of stations. Therefore, the brightness distribution $I(l,m)$ on the sky can be mapped by taking the Fourier transform of the visibilities \citep{Taylor1999} as given by
\begin{equation}
I_{\nu}(l, m) A_{\nu}(l,m) = \sum_k V_{\nu}(u_k, v_k) \, \rmn{e}^{i2\pi(u_kl+v_km)},
\end{equation}
where $A_{\nu}(l,m)$ is the primary beam response of the telescope, $l$ and $m$ are the direction cosines and the subscript $\nu$ indicates the frequency of the measurement. Additionally, each visibility contains a noise component $N_{\nu}(u_k, v_k)$. Therefore, the noise realization in the image plane $ n_{\nu}(l,m)$ is given by 
\begin{equation}
n_{\nu}(l,m) = \sum_k N_{\nu}(u_k, v_k) \rmn{e}^{i2\pi(u_kl+v_km)}.
\end{equation}

The sampling function $S_{\nu}(u, v)$ reflects the baseline distribution, and is given by
\begin{equation}
S_{\nu}(u, v) = \sum_k \delta^{2D} (u-u_k, v-v_k),
\end{equation}
where, $\delta^{2D}$ is the 2-dimensional Dirac delta function. We used uniform weighting after gridding visibilities on to the uv plane, i.e. all visibilities within a uv cell were averaged. Therefore, the root-mean-square (RMS) noise in the gridded uv plane is inversely proportional to $\sqrt{S(u,v)}$. 

In order to obtain realistic simulations of the noise, we filled the real and imaginary parts of the visibilities $N_{\nu}(u_k, v_k)$ with Gaussian random numbers. Visibilities were then Fourier transformed to the image space to obtain noise maps $n_{\nu}(l,m)$. By simulating the noise in this manner, we incorporated the realistic power spectrum of the noise into our simulations. The noise maps were normalized to have the appropriate RMS values. Based on the theoretical calculations of the system equivalent flux density \citep{Labropoulos2009}, we expect the RMS noise to be about 120 mK at the resolution of 3 arcmin (i.e. the full resolution offered by the LOFAR core), at 150 MHz, after 600 h and 0.5 MHz of integration for uniformly weighted
data. Fig. 4 shows the RMS noise used for normalizing the simulations as a function of frequency. The adopted noise values are indicative only, and they may change in the actual observations.

\begin{figure}
\centering
\includegraphics{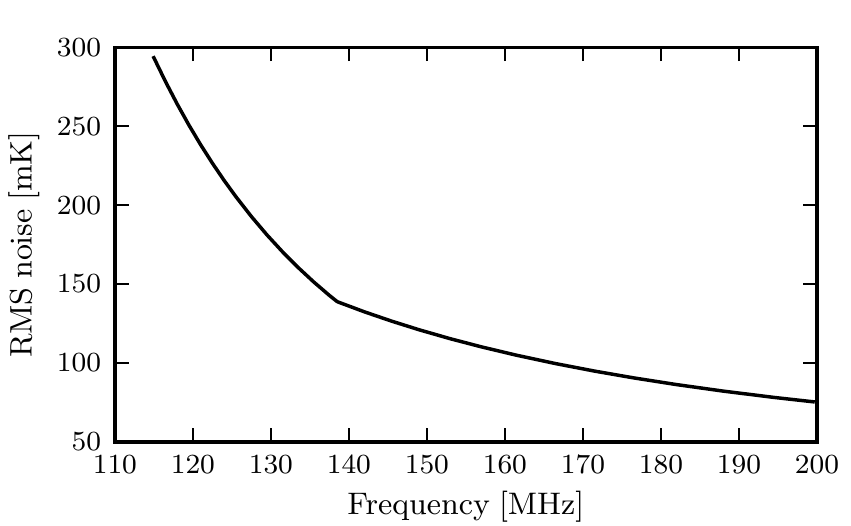}
\caption{Root Mean Square (RMS) system noise as a function of frequency after 600 h and 0.5 MHz integration, used for normalizing the noise simulations.}
\end{figure}

\subsection{uv-mask or PSF convolution}
An interferometer can only sample the spatial scales corresponding to its uv coverage. The effect of the uv coverage is equivalent to convolution with the point spread function (PSF) in the image plane. In order to mimic the effect of the PSF convolution, the simulated EoR, foreground and noise maps were Fourier transformed to the uv plane, multiplied by the uv coverage and Fourier transformed back to the image plane.

The uv coverage of an interferometer is frequency dependent because $u$ and $v$ are expressed in wavelengths. If a uv point is only sampled in a part of the bandwidth, it
could introduce discontinuities in the foregrounds and noise properties along the frequency dimension. This would affect the performance of the foreground removal algorithms which are based on the assumption that the foregrounds are spectrally smooth. In order to avoid such discontinuities, we maintained the same uv coverage throughout the bandwidth. This can be achieved by masking the intersections of the uv coverages at all frequencies \citep{Jelic2008, Bowman2009}. In other words, the uv points that were only partially covered in the bandwidth were discarded. Our uv mask allowed baselines between 40 and 800 wavelengths, assuming a complete uv coverage in this range. We only considered stations within the central core of LOFAR because these densely sample the corresponding part of the uv plane. Stations outside the core provide longer baselines and are used in actual observations to remove the point sources during the calibration. They would then be discarded in the subsequent analysis. The noise maps were simulated in the uv plane and already contain the uv coverage. However, they were also multiplied by the uv mask to maintain the same uv coverage at all frequencies.

In the case of real observations, independent gridding of visibilities at different frequencies can change the PSF by a small fraction at different frequencies. A chromatic PSF mixes the angular structures of foregrounds into the frequency direction, which has been dubbed as ``mode-mixing" in the literature \citep{Bowman2009, Datta2010, Vedantham2012, Morales2012, Hazelton2013}. Effects of uv gridding are not included in our simulations.

\subsection{Primary beam correction}
The image formed by an interferometer is the sky brightness distribution multiplied with its primary beam response as described by equation (8). Due to the primary beam response of the telescope, the strength of any observed signal from the sky (EoR and foregrounds) decreases away from the pointing direction. But noise, being uncorrelated among the visibilities, remains unaffected by the primary beam response. Hence the signal-to-noise (SNR) decreases away from the direction of pointing.

The primary beam response scales with wavelength. In the case of LOFAR, the Full Width Half Maximum (FWHM) of the primary beam changes from 4.75 degrees at 120 MHz to 2.85 degrees at 200 MHz \citep{vanHaarlem2013}. We find that the performance of the foreground removal suffers severely due to this frequency dependence, as shown further in Section 3.7. The primary beam correction reconstructs the frequency coherence of the foregrounds and hence improves the foreground removal. As a result of the correction, the EoR signal and the foregrounds have the same strength throughout the image but the noise increases towards the edges. Our simulations do not contain the primary beam response. Therefore, for a simple treatment of the primary beam, we consider as if the EoR and foreground simulations were already beam corrected, and we multiply only the noise maps by the reciprocal of the primary beam $1 / A_{\nu}(l,m)$ in the image space. We assume a Gaussian primary beam with the same FWHM as that of the measured response in \cite{vanHaarlem2013}.

In reality, the primary beam response resembles the $\mathrm{sinc}^2$ function and its correction requires division by zero around the nulls. However, a Gaussian is a good approximation of the primary beam within the first null and we restrict the image size to be well within the first null. One way to avoid the primary beam correction and still get desirable foreground removal, could be to maintain the same primary beam shape throughout the bandwidth. This could be achieved by convolving the visibilities with an appropriate kernel. However, it would restrict the field of view to the smallest possible case i.e. that obtained at the highest observation frequency. A better alternative would be to incorporate the beam model in the foreground removal algorithm. Our current efforts are focused on this front and we leave this topic for a future paper. 

We would like to note that some realistic issues are sidestepped due to our preliminary treatment of the primary beam. For instance, our simulations do not contain foreground sources in sidelobes of the primary beam, which may be an important source of the foreground contamination \citep{Yatawatta2013, Dillon2014}. We have also not considered the time and station-to-station variations of the beam.  More detailed modelling of the primary beam is required to study these effects, which we consider to be out of the scope of this paper.

\subsection{Gaussian smoothing}
The noise RMS depends on the resolution. The expected EoR signal RMS at the full resolution offered by the LOFAR core ($\sim$3 arcmin) is about 6 mK at 150 MHz, whereas the noise RMS is 120 mK after 600 h, 0.5 MHz integration. Therefore, the SNR at 3 arcmin resolution is very low. Not only may the signal detection be extremely difficult with such poor SNR, but the foreground removal may also be ineffective with such noisy data. The reason for poor SNR is the higher noise contribution at small spatial scales, which correspond to few long baselines. Even within the LOFAR core, the longer baselines are fewer in number, causing lower sampling density in the outer part of the uv coverage. Therefore the noise power is mostly concentrated on small spatial scales, as shown in Fig.5. We took advantage of this fact to reduce the noise significantly by smoothing the images with a Gaussian kernel, which is equivalent to multiplying the visibilities with a Gaussian. Therefore, by smoothing the images, we effectively down-weighted the longer baselines and reduced the noise. As shown in Fig. 6, the noise deceases rapidly with increasing smoothing scales up to few arcmin. For larger smoothing scales, the corresponding part of the uv plane is well sampled and therefore the rate of noise suppression decreases. The signal strength also decreases due to smoothing, but not as significantly as the noise (see Fig. 1). We find that the best SNR in the case of LOFAR is achieved when images are smoothed on a scale of 12 arcmin FWHM.

\begin{figure}
\centering
\includegraphics{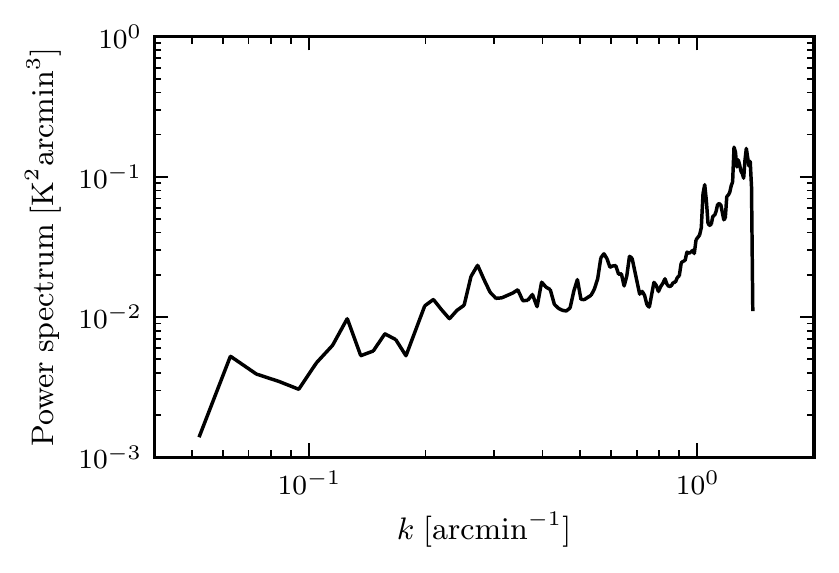}
\caption{Radially averaged 2-dimensional power spectrum of the noise at 150 MHz as a function of wavenumber $k = 2 \pi / \theta$, where $\theta$ is the angular scale. Most of the noise power is concentrated at large $k$ values due to the lower sampling density of the outer part of uv plane.}
\end{figure}

\begin{figure}
\centering
\includegraphics{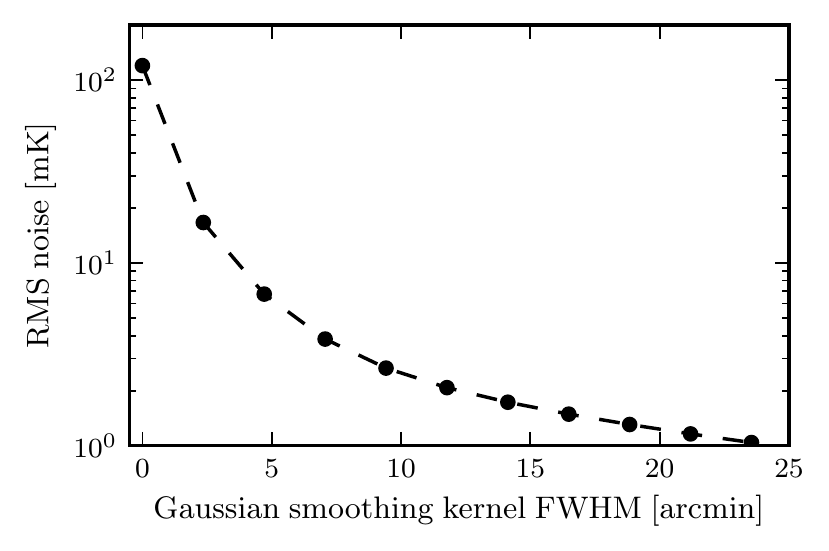}
\caption{RMS noise as a function of smoothing scale at 150 MHz after 600 h and 0.5 MHz integration. The smoothing scale of zero means no smoothing. The noise initially decreases rapidly with smoothing as we effectively down-weight the long baselines.}
\end{figure}

\subsection{Foreground removal}
The success of EoR experiments depends strongly on the accuracy of the foreground removal. As the foregrounds are 2 to 3 orders of magnitude larger than the EoR signal,
even small errors in their removal can severely affect the extraction of the underlying EoR signal. Foreground removal schemes are based on the assumption that the foregrounds are smooth along the frequency dimension, whereas the signal and noise are not \citep{Shaver1999,DiMatteo2002, Oh2003, Zaldarriaga2004}. The signal is not expected to be smooth in frequency because it varies in space. Below we briefly describe the three foreground removal methods used in this paper.

\begin{enumerate}{\setlength\itemindent{10pt}}
\item Generalized Morphological Component Analysis (GMCA): GMCA is a general source separation technique which utilizes morphological diversity and sparsity to identify different components in the data. The GMCA implementation of \cite{Chapman2013} finds a basis set in which spectrally smooth foreground components are sparsely represented and can hence be distinguished from the EoR signal and noise.

\item Wp smoothing: Wp smoothing \citep{Machler1995} was used by \cite{Harker2009} as an EoR foreground removal algorithm. It minimizes the sum of the squared difference between the foregrounds and the data, subject to a penalty on relative changes of curvature.

\item FASTICA: FASTICA is an independent component analysis technique and it was implemented by \cite{Chapman2012} as a foreground removal algorithm in the context of the EoR. It separates statistically independent components of the foregrounds by maximizing non-gaussianity of their mixture.

\end{enumerate}

\begin{figure}
\centering
\includegraphics{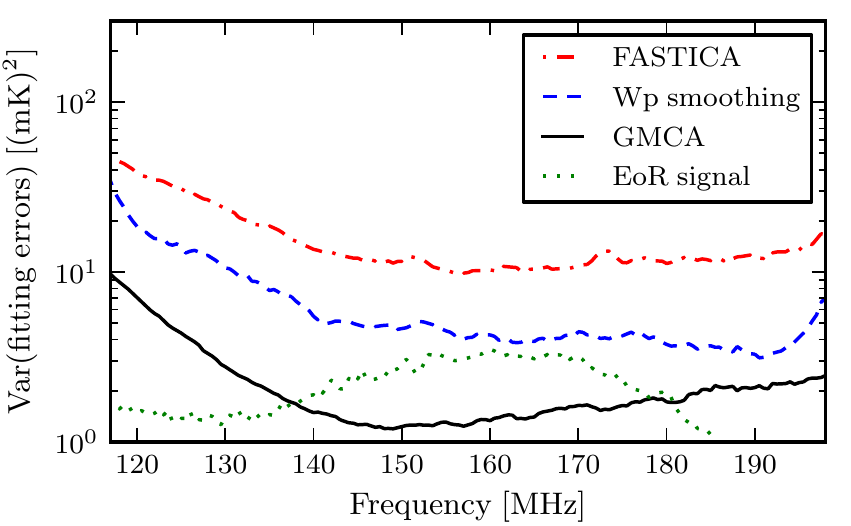}
\caption{Variances of the foreground fitting errors (i.e. the difference between originally simulated and reconstructed foregrounds) for three removal methods for 600 h and 0.5 MHz integration, compared with the EoR signal.}
\end{figure}

In Fig. 7, we compare the variance of the foreground fitting errors $\rmn{Var}(f - \hat{f})$ for the three removal methods, where $f$ and $\hat{f}$ are the originally simulated and reconstructed foregrounds, respectively. GMCA performs best among the three methods. We would like to note that further optimization might be possible for each of these methods. For the purpose of this paper, however, we choose GMCA to demonstrate the results.

We find that the primary beam correction improves the performance of the foreground removal. Fig. 8 shows that the foreground residuals are significantly reduced when the beam correction is applied. The EoR and foreground simulations were multiplied with the primary beam  in the image space in the case of no correction. The GMCA was run to find two independent components (see \cite{Chapman2013} for details). Due to the frequency dependent primary beam response, GMCA fails to capture the frequency coherence of the foregrounds when the beam correction is not applied. Instead, it tries to optimize the foreground residuals in two localized parts of the bandwidth as shown in Fig. 8.

\begin{figure}
\centering
\includegraphics{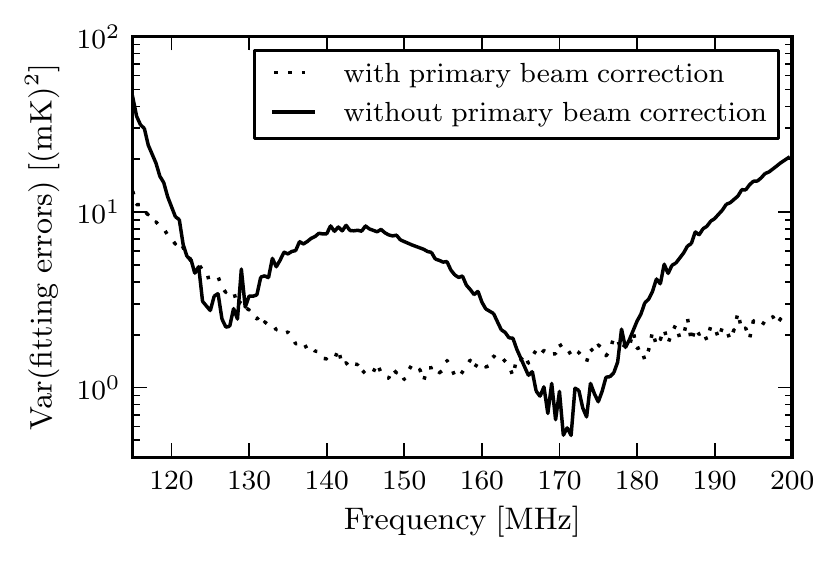}
\caption{Comparison of variances of the GMCA foreground fitting errors with and without primary beam correction for 600 h and 0.5 MHz integration. The beam correction reconstructs the frequency coherence of the foregrounds and hence improves the foreground fitting.}
\end{figure}

We have not considered removal of the polarized foregrounds separately, because the total intensity of the polarized foregrounds is smooth in frequency, these should be removed by the above algorithms. However, imperfect calibration of the instrumental polarization would lead to leakage of the polarized foregrounds into the total intensity. Such leakage would have frequency-dependent structure \citep{Jelic2010} and therefore it may not be removed by the above algorithms. We assume that the instrumental polarization will be well calibrated.

\section{Variance measurement and parameter estimation}

Our interest lies in extracting from the data the variance of the signal as a function of frequency. The variance of the data can be measured in the image plane as $\langle (X - \bar{X})^2 \rangle$ where $X$ is the flux density measured at a pixel, and $\bar{X}$ is the average flux density in the image. We measure the variance of the simulated data at every spectral channel in the image plane because the EoR signal and the foregrounds are simulated in the image plane. However, it is preferable to measure it in the uv plane for the actual observations because they are measured as visibilities.

The foreground extracted data $d$ contain the EoR signal $s$, the noise $n$ and foreground fitting errors $r$. The variance estimator can be expressed as
\begin{equation}
\rmn{Var} (d_{\nu}) = \langle d_{\nu}^2 \rangle = \langle (s_{\nu}+n_{\nu}+r_{\nu})^2 \rangle,
\end{equation}
where the subscript $\nu$ indicates the spectral channel. We have assumed in the above equation that the mean value of the data is zero, as is the case in interferometric images made with no zero spacing. The above equation can be further expanded as
\begin{eqnarray}
\rmn{Var} (d) &=& \langle s^2 + n^2 + r^2 + 2sr + 2nr \rangle		\nonumber \\
				 &=& \rmn{Var} (s) + \rmn{Var} (n) + \rmn{Var} (r) + 2 \langle rs+rn \rangle,
\end{eqnarray}
where the first equality follows because the signal and the noise are uncorrelated so their cross-correlation $\langle sn \rangle$ is zero. The subscript $\nu$ has been dropped for convenience but all quantities are measured for each spectral channel.

The signal variance $\rmn{Var}(s)$ can be estimated by measuring the variance of data $\rmn{Var}(d)$ and subtracting from it the expected noise variance $\rmn{Var}(n)$, foreground fitting error variance $\rmn{Var}(r)$ and the cross-correlation between the
noise and the foreground fitting errors $2 \langle rn \rangle$. (see equation
(12)). The noise and the foreground fitting errors are correlated due to the part of the noise that is removed by the foreground removal algorithm. We do not correct for the cross-correlation between the signal and the foreground fitting errors $2 \langle rs \rangle$ because it will not be possible to estimate it in the case of actual observations from the data. However, we believe it would not be very significant because we find from the simulations that the term $2 \langle rs \rangle$ is much smaller than other terms in equation (12). For the purpose of our simulations, variances of the noise, the foreground fitting errors and their
cross-correlation are estimated from many Monte Carlo realizations of the noise and the foregrounds. In the case of actual observations, such noise realizations will be obtained from the data by differencing consecutive spectral channels of very narrow bandwidths (12 kHz). The foregrounds and the EoR signal, being smooth on these scales, are expected to get subtracted. However, since the noise is uncorrelated in different spectral channels, channel differencing is expected to yield good estimates of the noise. 
Estimating the foreground fitting errors from the data might be difficult in the case of actual observations and we may have to rely on foreground simulations for this purpose. However, one possible way might be to split the data into two equal integration time intervals, run the foreground removal on the two data sets separately and then subtract the residuals of one from another. The signal will get subtracted in this operation and the noise estimate can be subtracted from the variance of the remaining maps to obtain the estimate of the fitting errors.

The statistical error in the variance estimation can be obtained by computing the variance of the variance estimator. For Gaussian noise, it is given by \citep{Casella2002}
\begin{equation}
Var(\hat{\sigma}^2) = \frac{2{\sigma}^4}{N},
\end{equation}
where $\hat{\sigma}^2$ is the variance estimator, $\sigma$ is the true noise RMS and $N$ is the number of measurement samples. Here, $\sigma$ is the noise RMS after the primary beam correction has been applied. As the noise increases away from the centre of the image, a larger image size implies a higher $\sigma$ and therefore a larger error. On the other hand, the larger the image size, the larger the number of independent samples $N$ for the variance measurement. Considering these two counteracting effects,
we determine the image size for variance measurement that minimizes the error. This leads to a frequency dependent image dimension, but takes advantage of the larger field of view at lower frequencies.

\subsection{Cross-variance measurement}

An alternative way to measure the variance is to cross-correlate consecutive spectral channels. We will refer to such measurement as `cross-variance'. We measure the cross-variance as $\langle X_i X_{i+1} \rangle$ where $X_i$ and $X_{i+1}$ are flux densities at the same pixel in $i^{\rmn{th}}$ and $(i+1)^{\rmn{th}}$ spectral channels, respectively. Measuring the cross-correlation of channels separated in frequency by $\Delta \nu$ is equivalent to measuring the power spectrum at a single scale parallel to the line of sight $k_{\parallel} = 2 \pi / \Delta\nu$. Therefore, the cross-variance gives the variance in the frequency direction on a single $k_{\parallel}$ mode. 

The cross-variance estimator can be expressed as
\begin{eqnarray}
\langle d_i d_{i+1} \rangle &=& \langle (s_i + n_i + r_i)(s_{i+1} + n_{i+1} + r_{i+1}) \rangle	\nonumber \\
&=& \langle s_i s_{i+1} + r_i r_{i+1} \rangle + 2 \langle s_{i}r_{i+1} + r_i n_{i+1} \rangle,
\end{eqnarray}
where we have assumed the pairs of cross terms such as $\langle s_i r_{i+1} \rangle$ and $\langle r_i s_{i+1} \rangle$ to be equal. The advantage of estimating the cross-variance as compared to the variance is that since noise is uncorrelated between different spectral channels, the cross term $\langle n_i n_{i+1} \rangle$ averages to zero. Similar to variance estimation, the signal and noise are uncorrelated, so their cross-correlation $\langle s_i n_{i+1} \rangle$ is zero. And the cross-correlation between
the signal and the foreground fitting errors $\langle s_i r_{i+1} \rangle$ is not corrected for. Therefore, the signal cross-variance $\langle s_i s_{i+1} \rangle$ is estimated by measuring the cross-variance of the data $\langle d_i d_{i+1} \rangle$ and subtracting from it the foreground fitting error cross-variance $\langle r_i r_{i+1} \rangle$ and the cross-variance of the noise and the foreground fitting errors $2 \langle r_i n_{i+1} \rangle$.

Unlike for variance estimation, the noise variance does not need to be subtracted by hand in the case of cross-variance estimation, reducing the chance for systematic
errors. Additionally, the statistical error in the measurement reduces by a factor of 2 in variance as we will show in equations (15) and (16). The variance of the cross-variance estimator $\hat{\sigma_{\rmn{c}}}^2$ is given by 
\begin{eqnarray}
\rmn{Var}(\hat{\sigma_{\rmn{c}}}^2) &=& \rmn{Var}[\rmn{E}(X_i X_{i+1})] = \rmn{E}[\rmn{Var}(X_i X_{i+1})] \nonumber \\
&=& \rmn{E} \left\lbrace \rmn{E}\left[ (X_i X_{i+1})^2 \right] - \left[ \rmn{E}(X_iX_{i+1}) \right]^2 \right\rbrace \nonumber \\
&=& \rmn{E} \left\lbrace \rmn{E}\left[ (X_i X_{i+1})^2 \right] \right\rbrace ,
\end{eqnarray}
where the last equality follows because the noise in two different spectral channels i.e. $X_i$ and $X_{i+1}$ is uncorrelated. For the same reason, equation (15) can be further simplified as 
\begin{equation}
\rmn{Var}(\hat{\sigma_{\rmn{c}}}^2) = \rmn{E} \left[ \rmn{E}(X_i^2) \rmn{E}(X_{i+1}^2) \right] = \frac{{\sigma_{i}}^2 {\sigma_{i+1}}^2}{N},
\end{equation}
where $\sigma_{i}$ and $\sigma_{i+1}$ are the RMS noise in the $i^{\rmn{th}}$ and $(i+1)^{\rmn{th}}$ spectral channels respectively.

The cross-variance of the signal is slightly lower than its variance at 12 arcmin (FWHM), 0.5 MHz resolution scale. This is because small-scale structures which are coherent on scales smaller than 1 MHz do not contribute to the cross-variance measurement. However, the aforementioned advantages of cross-variance estimation supersede this disadvantage, as we will show in Section 5. Cross-correlation of two sub-epochs of the observation period has similar advantages, and it has been considered by \cite{Harker2010} in the context of power spectrum estimation.

We realized in hindsight that the cross-correlation of two frequency channels in order to detect the EoR signal has been previously proposed by \cite{Bharadwaj2001}.

\subsection{Parameter estimation}
Once the signal variance has been extracted, we fit the model described in Section 2 to it and estimate the best-fitting parameters. We used the Markov Chain Monte Carlo (MCMC) technique to explore our 4D parameter space - $z_r, \Delta z, \beta, A$. MCMC maps the posterior probability distribution $P(\theta | D)$ of the model parameters $\theta$ given the observed (here simulated) data $D$. The best-fitting parameters are obtained at the point in parameter space where the posterior is maximized. The posterior can be obtained from the likelihood $P(D | \theta)$ and the prior $P(D)$ by Bayes' theorem:
\begin{equation}
P(\theta | D) \propto P(D | \theta) P(D).
\end{equation}
We assume uniform priors and therefore mapping the posterior is the same as sampling the likelihood.  We used the code \textsc{emcee} \citep{Foreman-Mackey2013} to map the likelihood and ultimately find the maxima. The code uses multiple random walkers to sample the likelihood function. At each step, the likelihood is computed assuming Gaussian noise as
\begin{equation}
P(D | \theta) = \prod_{i} \frac{1}{\sqrt{2 \pi \sigma_{n,i}^2}} \exp{ \frac{-(D_i - M_i(\theta))^2}{2 \sigma_{n,i}^2}},
\end{equation}
where $D_i$, $M_i(\theta)$ are the measured and predicted (by model) variance values for the $i^{\rmn{th}}$ spectral channel, and $\sigma_{n,i}^2$ is the variance of the error in measurement as given by equation (13) or (16). 

In order to sample the parameter space, \textsc{emcee} iteratively draws samples for each random walker using a proposal distribution based on the current positions of other walkers. If the likelihood at the proposed position is higher than the current one, the step is accepted. If it is lower than the current likelihood, it is accepted with a certain probability. Over time, the chain explores the parameter space and maps the likelihood function. To obtain the posterior distribution of reionization parameters, we marginalize over the remaining model parameters i.e. the scaling amplitude $A$ and the power law index $\beta$ as
\begin{equation}
P(z_\rmn{r}, \Delta z | D) = \int P(z_\rmn{r}, \Delta z, A, \beta | D) \; \rmn dA \; \rmn d \beta.
\end{equation}

In principle, we should use the chi-squared distribution for the likelihood function, because the error on the variance measurement of a Gaussian noise follows a chi-square distribution. However, we get almost the same results for both Gaussian and chi-squared distributions. This is the case because a chi-square distribution with large degrees of freedom converges to a Gaussian distribution, suggesting that the number of samples in our measurement suffices for the central limit theorem to hold. 

\section{Results}
The simulation pipeline described in Section 3 generates the mock observational data sets. We estimate the signal variance from the mock data and fit the model described in Section 2 to it. Fig. 9 shows results of the model fitting and parameter estimation for a realization of the data. The top panel shows the actual and estimated signal variance and the best model fit to the data for 600 h of integration. Foreground removal and variance measurement are performed at 0.5 MHz resolution but the variance measurements are then averaged to 5 MHz. Such re-binning of the data is done only for the convenience of showing the results but it does not affect the model fitting. The horizontal dotted line of zero variance is drawn to illustrate the significance of the detection. The bottom panel shows the obtained marginalized posterior probability densities of the reionization parameters. The actual values of the used EoR simulation parameters were $z_r = 7.68$ and $\Delta z = 0.43$. The extracted parameter values are $z_r = 7.72^{+0.37}_{-0.18}$ and $\Delta z = 0.53^{+0.12}_{-0.23}$. The errors are given at 68 per cent confidence.

\begin{figure}
\centering
\includegraphics[]{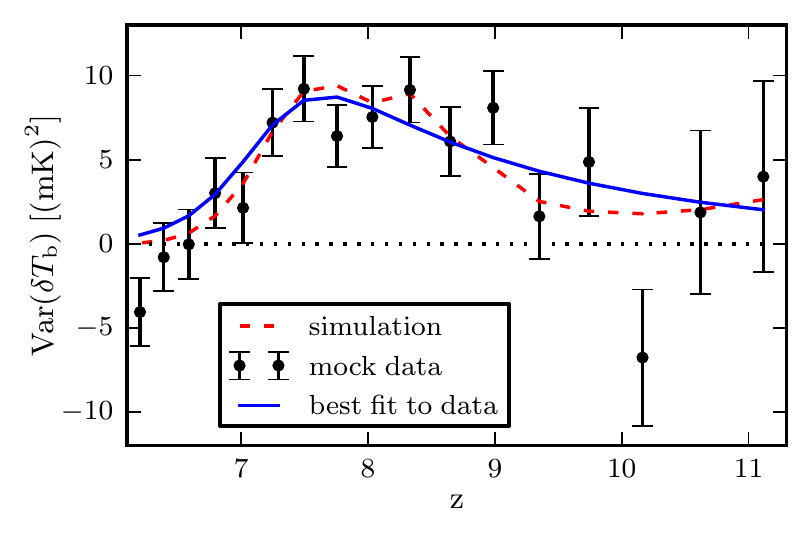}
\includegraphics[]{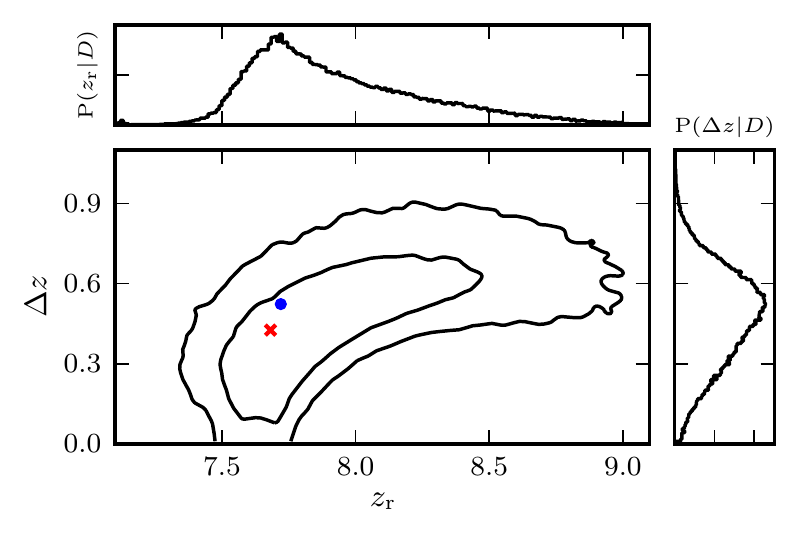}
\caption{Model fitting and parameter estimation results for a mock data set of 600 h of integration. Top panel: Variance of $\delta T_{\rmn{b}}$ as a function of redshift. Dashed curve: original EoR signal; black points: estimated variance; solid curve: model fit to the data. Error bars are calculated from equation (13). Bottom panel: constraints on the reionization parameters with 68 and 95 per cent confidence levels. The cross and dot show the actual and best-fitting values of the parameters respectively. Marginalized probability densities are plotted on the sides.}
\end{figure}

Fig. 10 shows the marginalized probability density functions (PDF) for the scaling amplitude $A$ and the power law index $\beta$. Assuming the null hypothesis to be $A=0$, $A$ rises to a significance of 4 standard deviations in 600 hours. We would call such measurement as the detection of the signal with a significance of 4 standard deviations. However, we would like to note that $A$ could rise to a high significance level due to systematic errors in the case of actual observations. Therefore, it is important to be able to extract not only $A$, but also $z_r$ and $\Delta z$ within desirable ranges in order to claim a detection of the signal.

\begin{figure}
\centering
\includegraphics[]{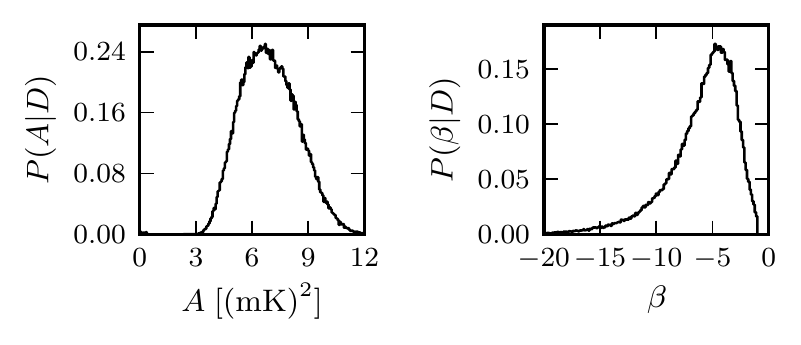}
\caption{Marginalized probability densities for the scaling amplitude $A$ and the power-law index $\beta$. $A$ rises to a significance of 4 standard deviations above the null hypothesis in 600 h of integration.}
\end{figure}

It is clear from Fig. 10 that the power law index $\beta$ is not constrained well with 600 h of integration. This is due to the poor sensitivity and the limitation of the model at high redshifts. However, the uncertainty in  $\beta$ does not significantly affect the variance during the peak of reionization and hence it does not affect the estimation of the EoR parameters $z_\rmn{r}$ and $\Delta z$.

The results of the cross-variance extraction are shown in Fig. 11. Posterior probability distributions of all model parameters and their pairs are shown in Fig. 12. The statistical uncertainty in the cross-variance measurement is lower than that of the variance, as shown in equations (15) and (16). Therefore, the significance of detection and constraints on the model parameters improve in the case of the cross-variance. The scaling amplitude $A$ rises to a significance of 7 standard deviations, and the extracted values of the EoR parameters improve as: $z_r = 7.73^{+0.20}_{-0.16}$ and $\Delta z = 0.44^{+0.10}_{-0.09}$. However, we would like to note that the results in this paper are based on the assumption that many systematic errors would remain under control. These include calibration errors, foreground contamination due to sources in the sidelobes of the primary beam, effects of uv gridding, ionosphere, etc.

\begin{figure}
\centering
\includegraphics[]{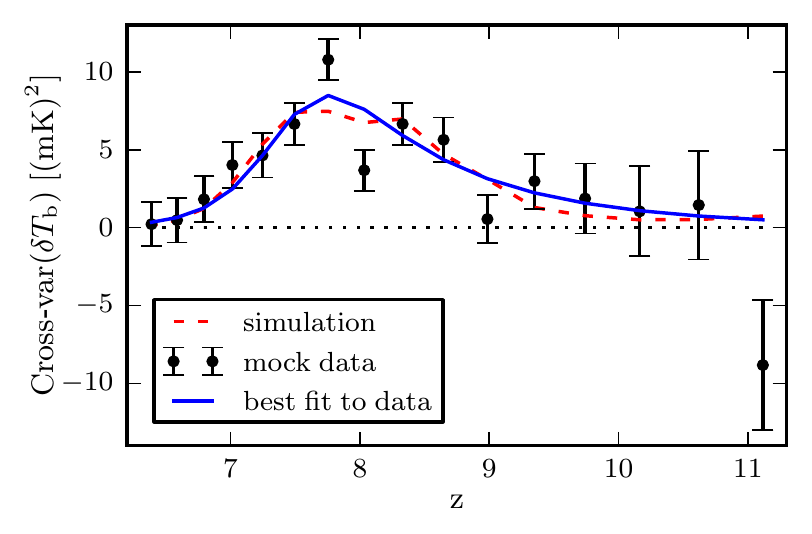}
\caption{Cross-variance extraction after 600 h of integration, obtained by cross-correlating adjacent frequency channels. The conventions are same as in Fig. 9 except that the error bars are calculated from equation (16). The constraints on all model parameters are show in Fig. 12. The cross-variance measurement improves the results because noise is uncorrelated in different frequency channels.}
\end{figure}

\begin{figure*}
\centering
\includegraphics[]{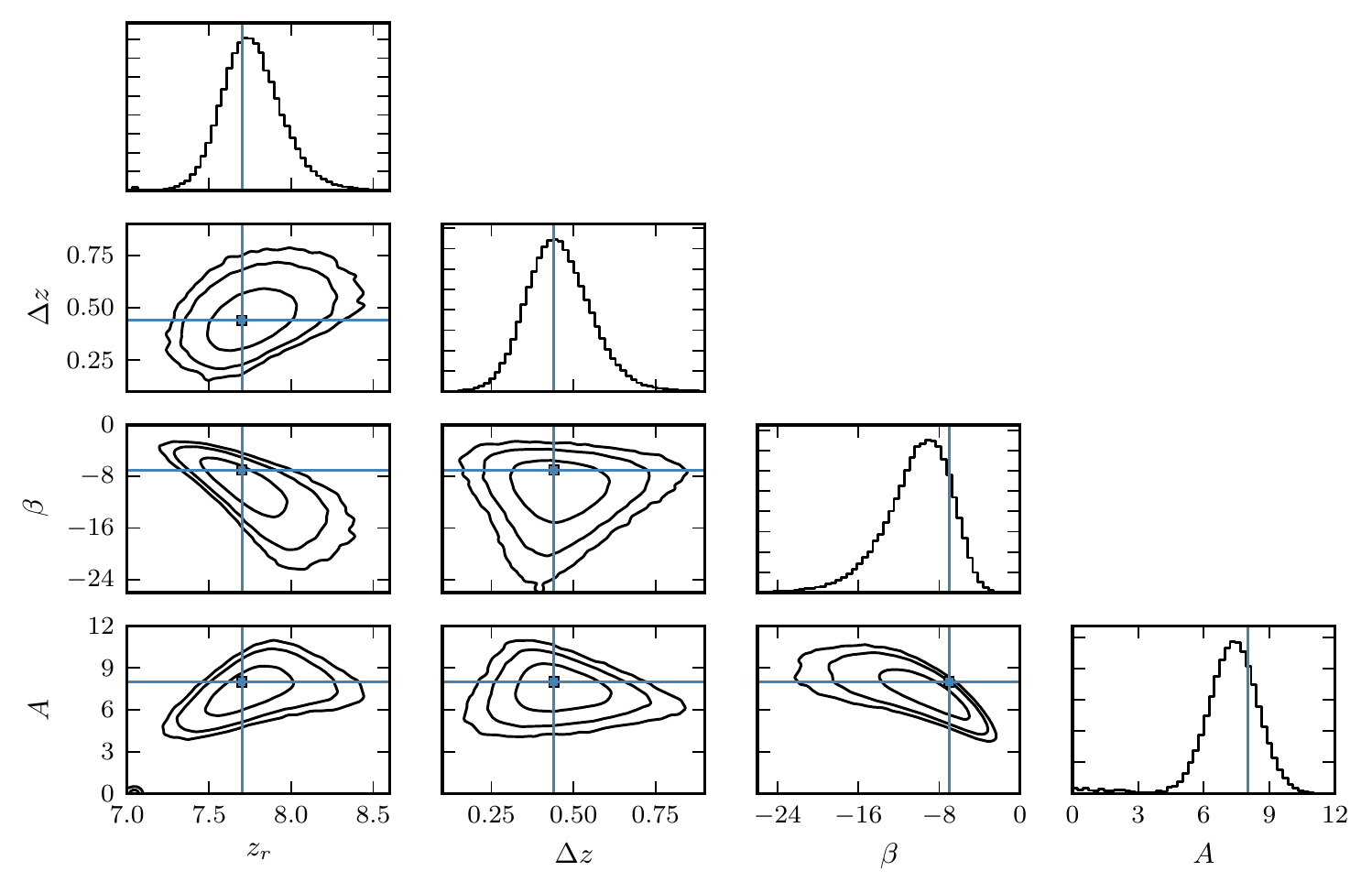}
\caption{Posterior probability distributions of the model parameters obtained from the cross-variance measurement after 600 h of integration. The parameters are the reionization redshift ($z_r$), duration of reionization ($\Delta z$), power law index ($\beta$) and the scaling amplitude ($A$). The contours show 68.3, 95.4 and 99.7 percent confidence levels, whereas the horizontal or vertical lines mark the true values of the parameters. The significance of detection rises to 7 standard deviations, and constraints on parameters improve with the cross-variance measurement as compared to the variance.}
\end{figure*}

A measurement of the variance of the signal can start to give some insights about the physics of reionization. The signal variance can be used to distinguish between inside-out and outside-in models of reionization \citep{Watkinson2013}. Measuring the redshift and duration of reionization will provide important constraints for the simulations and theoretical models, which then can improve our understanding of the EoR and the first objects in the Universe.

\subsection{A different reionization history}
The results shown in Fig. 9, 11 and 12 indicate that LOFAR can in principle detect and constrain the EoR for a particular history of reionization. However, the exact redshift of reionization is unknown. Therefore, it is necessary to test whether it would be possible to constrain the EoR parameters for a different history of reionization. In particular, the signal detection may become more difficult if reionization was completed at higher redshifts because the system noise increases at lower frequencies. We therefore perform the same exercise of parameter estimation for another simulation with $z_r = 9.30$ and $\Delta z = 0.61$. The results are shown in Fig. 13. The extracted values of the parameters are: $z_r = 9.60^{+0.41}_{-0.37}$ and $\Delta z = 0.55^{+0.21}_{-0.13}$. Although the amplitude $A$ remains at a significance of 4 standard deviations, for the same quality of the data (i.e. 600 h and 0.5 MHz integration), we obtain weaker constraints on the EoR parameters.

\begin{figure}
\centering
\includegraphics[]{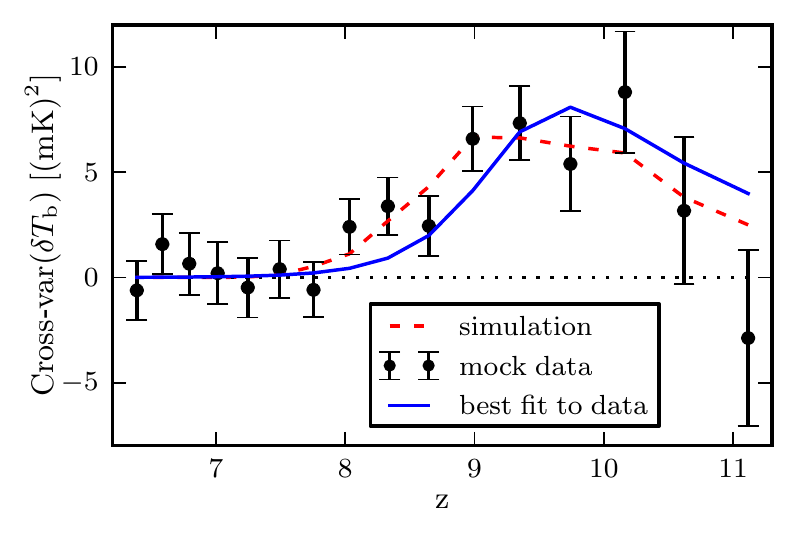}
\includegraphics[]{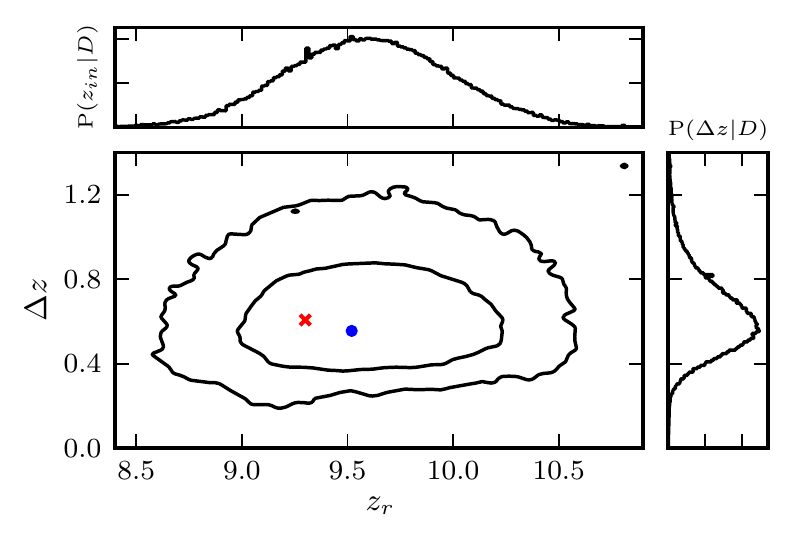}
\includegraphics[]{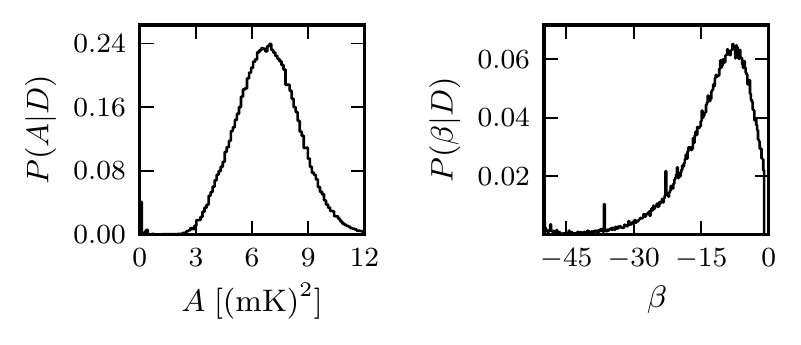}
\caption{Cross-variance extraction for a different history of the EoR, where reionization was completed earlier. Due to the higher system noise at low frequencies, the errors on the estimated parameters become larger if reionization were to occur at higher redshifts. The bottom panel shows the marginalized posterior probability densities of parameters $A$ and $\beta$.}
\end{figure}

\subsection{Better quality data}
The LOFAR-EoR project plans to acquire a few thousand hours of data over the coming years in order to constrain and understand the process of reionization. Therefore, it is important to  check whether the proposed model for the variance statistic with its limitations would work for better quality (higher SNR) data in the future with LOFAR or the Square Kilometer Array (SKA) \citep{Mellema2013}. For this purpose, we simulated observational data sets for 1200 h of integration, keeping all other parameters same the as before. Fig. 14 shows the results of the variance measurement for such a data set. The extracted values of the parameters are: $z_r = 7.71^{+0.13}_{-0.11}$ and $\Delta z = 0.44^{+0.07}_{-0.09}$. As expected, the constraints on the reionization parameters are improved because of the reduced noise. However, some of the systematic errors become significant, which we otherwise neglected for the 600-hour case. The signal variance is under-estimated. This is due to the part of the signal that is removed by the foreground removal algorithm. We  ignored the correlation between the signal and the foreground fitting errors, but with 1200 h of integration, the error bars become small enough for the effect of this correlation to become noticeable. The systematic errors introduced do not bias the estimates of the reionization parameters, as shown in the bottom panel of Fig. 13. However, the underestimation of the signal variance biases the scaling amplitude $A$ to lower values. The degeneracy between $A$ and the foreground removal can partially be broken by using multiple foreground removal methods. The bias in estimation of $A$ is a minor issue, but it can not be completely solved as any currently available foreground removal method would remove a small fraction of the EoR signal. Therefore, for a precise extraction of the signal, the foreground removal algorithms would need to be improved. 

The statistical errors at higher redshifts, where the model does not describe the signal well, remain too large to affect the fitting. But it can be seen that with $\sim$2000 h of integration, the mismatch between the model and the signal at the beginning of reionization would become a source of systematic errors. We hope that the development in theory and simulations over the coming years will enable us to use improved models to describe the history of reionization. 

\begin{figure}
\centering
\includegraphics[]{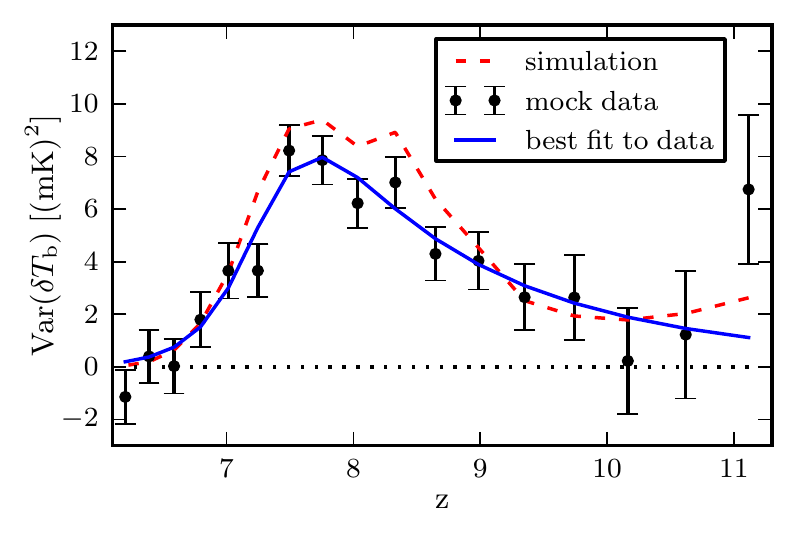}
\includegraphics[]{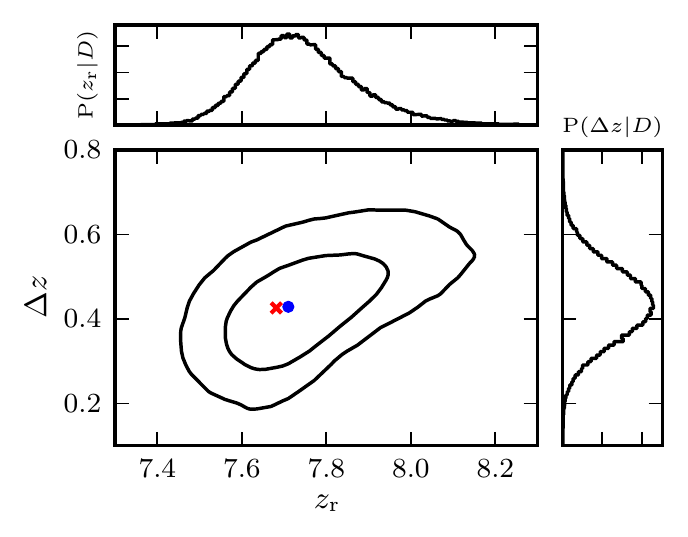}
\caption{Variance extraction and parameter estimation results for 1200 h of integration. The constraints on reionization parameters improve with better signal-to-noise ratio, but the systematic error due to the part of the signal that is removed by foreground removal starts to become significant in the model fitting.}
\end{figure}

\section{Conclusions}
We have investigated the extraction of the variance of the redshifted 21-cm emission as a tool to detect and constrain the global history of reionization. We have used simulations of the LOFAR-EoR case to demonstrate that the variance measurement is a promising tool for EoR experiments. 

We parametrized the evolution of the EoR signal variance with redshift in terms of four model parameters including a characteristic redshift and duration of reionzation. We then generated mock observations using a simulation pipeline to test the variance statistics. This study has helped us to realise the implications of instrumental characteristics such as uv coverage and primary beam response on the observations, and to investigate different strategies for data analysis. For instance, we have quantified the impact of the chromatic primary beam on the foreground removal to a first order.

We have shown that LOFAR should be able to detect the EoR signal with a significance of 4 standard deviations in 600 h of integration on a single field, assuming that the calibration errors are small and the point sources can be adequately removed. Additionally, it should be able to constrain the redshift $z_r$ and duration $\Delta z$ of reionization. We used a simulation with $z_r = 7.68$ and $\Delta z = 0.43$ to test our parameter estimation pipeline, and have been able to extract the parameters as $z_r = 7.72^{+0.37}_{-0.18}$ and $\Delta z = 0.53^{+0.12}_{-0.23}$, where the error bars are given at 68 per cent confidence. For the same quality of the data, we should be able to get better constraints by measuring the cross-variance i.e. the cross-correlation between consecutive spectral channels. The cross-variance improves the results because the noise in different spectral channels is uncorrelated. The cross-variance measurement can enable us to detect the signal with a significance of 7 standard deviations, and extract the EoR parameters as $z_r = 7.73^{+0.20}_{-0.16}$ and $\Delta z = 0.44^{+0.10}_{-0.09}$ for the same 600 hours of data.

If reionization was completed at higher redshifts where the sensitivity of LOFAR decreases, the constraints on the EoR parameters would be poorer. For a simulation with $z_r = 9.30$ and $\Delta z = 0.61$, we have been able to constrain the reionization parameters as $z_r = 9.60^{+0.41}_{-0.37}$ and $\Delta z = 0.55^{+0.21}_{-0.13}$. 

The constraints on the EoR parameters will improve with more hours of integration. For 1200 h of integration and a simulation with $z_r = 7.68$ and $\Delta z = 0.43$, we have been able to constrain the EoR parameters as $z_r = 7.71^{+0.13}_{-0.11}$ and $\Delta z = 0.44^{+0.07}_{-0.09}$. However, the systematic errors due to the part of the signal that is removed by the foreground removal algorithm will become significant with better quality data. Therefore, the foreground removal algorithms and the models of reionization would then need improvements for an accurate extraction of the EoR signal. 

Many realistic effects are not included in our analysis due to limitations of our simulations. These include calibration errors, foreground contamination due to sources in sidelobes of the primary beam, effects of uv gridding, ionosphere, RFI flagging and the polarization leakage. Our results are based on the assumption that these issues can be adequately controlled. The future work could focus on addressing these issues in order to optimize the data analysis strategy. Eventually, the goal is to apply the variance statistic to LOFAR-EoR observations.

\section{Acknowldegement}
AHP and SZ would like to thank the Lady Davis Foundation and The Netherlands Organization for Scientific Research (NWO) VICI grant for the financial support. VJ acknowledges the NWO for the financial support under VENI grant - 639.041.336. FBA acknowledges the support of the Royal Society via a University Research Fellowship. LVEK, AG, HKV, KMA and SD acknowledge the financial support from the European Research Council under ERC-Starting Grant FIRSTLIGHT - 258942.

\bibliographystyle{mn2e}
\bibliography{patil.bib}

\end{document}